\documentstyle[12pt,a4,epsfig]{article}

\def\nextline{\hfill\break}

\def\mycomm#1{\nextline\strut\kern-3em{\tt ====> #1}\nextline}

\def\gray{\special{ps: 0.4 setgray}}
\def\black{\special{ps: 0.0 setgray}}

\newcommand{\draft}{
\newcount\timecount
\newcount\hours \newcount\minutes  \newcount\temp \newcount\pmhours
 
\hours = \time
\divide\hours by 60
\temp = \hours
\multiply\temp by 60
\minutes = \time
\advance\minutes by -\temp
\def\hour{\the\hours}
\def\minute{\ifnum\minutes<10 0\the\minutes
            \else\the\minutes\fi}
\def\clock{
\ifnum\hours=0 12:\minute\ AM
\else\ifnum\hours<12 \hour:\minute\ AM
      \else\ifnum\hours=12 12:\minute\ PM
            \else\ifnum\hours>12
                 \pmhours=\hours
                 \advance\pmhours by -12
                 \the\pmhours:\minute\ PM
                 \fi
            \fi
      \fi
\fi
}
\def\fullclock{\hour:\minute}
\begin{centering}
\gray
\special{ps: -90 rotate}
\special{ps: -4600 -5100 translate}
\font\Hugett  =cmtt12 scaled\magstep4
{\Hugett Draft: \today, \clock}
\black
\special{ps: 90 rotate}
\special{ps: 5100 -4600 translate}
\end{centering}
\vskip -1.7cm
$\phantom{a}$
} 

\newcommand{\bmath}{\begin{displaymath}}
\newcommand{\emath}{\end{displaymath}}
 
\def\beq{\begin{equation}}
\def\eeq{\end{equation}}
 
\newcommand{\bea}{\begin{eqnarray}}
\newcommand{\eea}{\end{eqnarray}}
 
\def\eqref#1{(\ref{#1})}

\newcounter{saveeqn}

\catcode`\@=11 

\def\eqarraylabel#1{\@bsphack \if@filesw 
{\let \thepage \relax \def \protect {\noexpand \noexpand \noexpand }%
\edef \@tempa {\write \@auxout {\string \newlabel 
{#1}{{\mbox{\arabic{saveeqn}}}{\thepage }}}}\expandafter }\@tempa 
\if@nobreak \ifvmode \nobreak \fi \fi \fi \@esphack}

\catcode`\@=12 
 
\newcommand{\la}{$\Lambda$ }
\newcommand{\lab}{$\bar{\Lambda}$ }

\def\tsize{\Large} 
\def\asize{\normalsize} 
\title {
\begin{flushright}
\normalsize TAUP - 2593-99\\
\normalsize WIS99/30/Aug.-DPP
\end{flushright}
\vspace{2.0cm}
\tsize
Expected Polarization of $\Lambda$ particles produced in deep \\
\tsize
inelastic polarized lepton scattering\thanks{Supported
in part by grants from US-Israel Bi-National Science Foundation
and from the Israeli Science Foundation.}
}
\author{\asize
D. Ashery$^{1}$ \\
\asize and\\
\asize $\phantom{a}$ H. J. Lipkin$\,^{1,2}$
\vspace{0.5cm}
\\
\asize \sl $^1$\,School of Physics and Astronomy\\
\asize \sl The Raymond and Beverly Sackler Faculty of Exact Sciences\\
\asize \sl Tel Aviv University, 69978 Tel Aviv, Israel\\
\asize \sl and\\
\asize \sl $^2$\,Department of Particle Physics\\
\asize \sl The Weizmann Institute of Science, 76100 Rehovot, Israel\\
}
\date { }
\begin {document}
\maketitle
\begin{abstract}
\noindent
We calculate the polarization of \la and \lab particles produced in deep 
inelastic polarized lepton scattering. We use two models: the na\"{\i}ve 
quark model and a model in which SU(3)$_F$ symmetry is used to deduce
the spin structure of SU(3) octet hyperons from that of the proton. We
perform the calculations for \la and \lab produced directly or as decay
products of $\Sigma^0$ and $\Sigma^*$.
  \end{abstract}
\thispagestyle{empty} 
\draft
\newpage

The spin structure function of the nucleon has been studied extensively 
during the past several years. Very different experimental systems utilizing
polarized muon \cite{smc}, electron \cite{slac} and positron \cite{herm}
beams and covering different kinematic regions obtained results which are
consistent with each other. This effort led to precise measurements of the 
nucleon spin structure functions $g_1^N(x)$ (N = n,p) and their integrals
$\Gamma_1^N = \int{g_1^N(x)dx}$. 
Here $x \equiv  x_{Bj} = Q^2/2M\nu$ is the Bjorken scaling variable, $-Q^2$
is the four-momentum transfer to the target nucleon squared, $\nu$ is
the energy loss of the lepton in the laboratory frame, and $M$ is the
nucleon mass.
The interpretation of these results is 
that quarks in the 
nucleon carry only $\sim$30\% of the nucleon spin and that the strange
(and non-strange) sea is polarized opposite to the polarization of the valence
quarks. This interpretation leads to the question of where is the rest of
the nucleon spin is coming from which will not be addressed here.
Other open questions are: 1) how sensitive
are the conclusions to the SU(3) symmetry assumed in the process of
interpretation of the experimental data? 2) what is the mechanism that 
polarizes the strange sea? 3) how well do we understand the spin structure
of other hadrons? 4) are the quarks and antiquarks in the sea equally
polarized? \\

These questions may be addressed through measurement of the polarization
of \la and \lab produced in polarized lepton Deep Inelastic Scattering
(pDIS) on unpolarized targets. Having the strange quark as valence rather
than sea quark, the
polarization of \la particles is sensitive to its contribution to the
\la spin. By measurements of both \la and \lab polarisation both
$s$ and $\bar{s}$ roles can be studied. Recently some experimental
results became available \cite{e665} and more are expected in the
future \cite{herm,comp}.
As an illustration, a simple mechanism that can produce \la polarization
in pDIS is when the polarized virtual photon is absorbed by a strange
quark in the target nucleon sea. The struck quark emerges with helicity in
the direction of that of the photon. When it hadronizes into a \la (in the
current fragmentation region), it is likely to become a valence quark
in the \la and the
na\"{\i}ve quark model predicts that the polarization of the \la will be
the same as that of the strange quark. 
An experiment that can probe the low $x_{Bj}$ region will have a significant
fraction of the \la and \lab originating from this mechanism and will have 
sensitivity to the strange sea in the target nucleon. 
The transfer of polarization from the beam to the produced \la and \lab
is controlled by their helicity difference fragmentation functions
$\Delta \hat{q}$ \cite{jaf}. 
In recent measurements at LEP $\Delta \hat{q}$ was measured from \la 
polarization near the Z pole and was found to be consistent with the
na\"{\i}ve quark model expectations \cite{aleph,opal}. It is not unreasonable,
therefore, to expect that in the current fragmentation region of pDIS the
polarization will follow the same expectations. An alternative approach
is presented in \cite{jaf} where SU(3)$_F$ symmetry is assumed and
the spin structure of SU(3) octet hyperons is deduced from that of the proton.
Other mechanisms exist and are discussed in
\cite{def,thom,ekks,elliskk}.\\


A difficulty present in most experiments is that they cannot distinguish 
between \la and \lab
produced directly from hadronization of a struck quark or as decay
products. The main contributions from decays are the
$\Sigma^0 \rightarrow \Lambda \gamma$ and $\Sigma^* \rightarrow
\Lambda\pi$. The purpose of the present work is to calculate the expected
polarization of \la and \lab produced directly or as decay products. The
relative ammount of each component depends on the experimental conditions.
We carry out the calculations using two models.
One is the na\"{\i}ve quark model where all baryons are three-quark 
states with wave functions having zero orbital angular 
momentum and all the spin of the baryon comes from quark spins. In
the other, which we will refer to as 
the SU(3)$_F$ symmetry model, the spin structure of SU(3) octet hyperons
can be deduced from that of the proton \cite{jaf}. As we show, introduction 
of SU(3)$_F$ symmetry breaking does not affect the calulated \la polarization.
In both models we calculate the polarization of \la and \lab produced
directly or as decay products. Earlier calculations ignored the
\la and \lab produced as decay products except for \cite{koz} where
their polarization is calculated for the na\"{\i}ve quark model only.\\

    The inclusion of the contribution of $\Lambda's$ produced in $\Sigma^*$
decay has been questioned because the quark-gluon fragmentation process should
already include the production and decay via strong interactions of the
$\Sigma^*$. Clearly the $\Sigma^*$ intermediate state must be already included
in any fragmentation function which takes into account $all$ strong
interactions in the description of a process in which a struck quark turns into
a $\Lambda$ plus anything else.  But a precise value for such a fragmentation
function must also include the precise value of small quantitites like the
$\Lambda-\Sigma$ mass difference, calculated from first principles, in order to
take into account the effect of this mass difference on the relative branching
ratios of the $\Sigma^*$ into different final states. Such fragmentation
functions are not available at this time and there is some question whether
they will ever be realistically available. 

    The existing Monte Carlo programs for producing such fragmentation functions
do not go into such details and rely instead on a number of free parameters 
which are adjusted to fit vast quantities of data. These give separately the
numbers of $\Lambda$'s produced directly and those via decay of $\Sigma^*$'s.
We therefore use these Monte Carlo calculations and consider the two 
contributions 
separately. The $\Sigma^o$ decays electromagnetically. Its decay is never 
included in any strong interaction fragmentation function and the $\Lambda's$
produced via its production and decay must be considered separately in all
fragmentation models.  We note that the measured results of \la polarization
at the Z pole \cite{aleph} were reproduced by Monte Carlo simulations which 
included a
contribution from $\Sigma^*$ decay of about 8\% out of a total of 39\%. \\

We are interested in describing the process in which a polarized struck quark
$q$ with spin projection $m=+1/2$ picks up two quarks with two
other flavors and fragments into a baryon denoted by $B$ with spin projection M
and valence quark constituents (uds). The baryon $B$ is either a $\Lambda$, or
it is a $\Sigma^o$ or $\Sigma^*$ which decays into a $\Lambda$. We wish to
calculate the polarization of this $\Lambda$. Antiquarks and antibaryons
are handled in the same way.

Our basic assumptions are:

1. The polarization of the struck quark is maintained during the fragmentation
process.

2. All polarization states of the addtional two quarks are equally probable.

3. The relative amounts of $\Lambda$, $\Sigma^o$ and $\Sigma^*$ are changed
during the fragmentation process. They cannot be calculated from theory, but 
are determined by some Monte Carlo model of the fragmentation process
\cite{e665,aleph,opal,koz,gus}.

4. The relative amounts of the different $M$ states of the same baryon are
maintained during the fragmentation process, which is rotationally invariant
and is not correlated with the direction of the incdent beam.

5. The measured $\Lambda$ polarization is obtained by averaging the polarization
obtained for a given process over all the $M$ states relevant to this process.

In performing calculations with angular momentum algebra and Clebsch-Gordan
coefficients, one must note that the coefficients for a given $M$ state give
the relative probabilities for producing a $\Lambda$, $\Sigma^o$ or
$\Sigma^*$ in this $M$ state and the sum of these probabilities is normalized
to unity. However, these relative probabilities are not relevant to the 
experimental results because of the differences in fragmentation.
The relative probabilities of producing different $M$ states of the SAME
baryon can be calculated using these same Clebsch-Gordan coefficients, because
the different $M$ states of the extra two quarks are equally populated. However
these relative probabilities are not normalized to unity and must be explicitly
normalized.
 
With these asumptions the calculation of the $\Lambda$ polarization is 
straightforward if we know the wave function of the $\Lambda$ or have 
suffcient spin information on its spin structure. In the simple quark model
the spin wave function is given. However, we know from deep inelastic scattering
that the simple quark model does not adequately explain the spin structure of
the proton. The question then arises how  information  on the spin
structure of the $\Lambda$ can enable us to go beyond the simple quark model.

We first consider the probabilities that a struck quark 
$q$ with spin up, denoted by $q\uparrow$, will fragment into a
spin-1/2 Baryon with spin up or spin down, denoted respectively 
by $B\uparrow$ and $B\downarrow$. The ratio of these probabilities under the 
above assumptions is
\beq 
{{W(q\uparrow \rightarrow B\uparrow)} \over 
{W(q\uparrow \rightarrow B\downarrow)}}  =
{{q\uparrow (B\uparrow)} \over {q\uparrow (B\downarrow)}}  =
{{q\uparrow (B\uparrow)} \over {q\downarrow (B\uparrow)}}  
\label{WW1}
\eeq
where $q\uparrow (B\uparrow)$ and $q\uparrow (B\downarrow)$ denote the
distribution of quark $q$ with spin up  in a baryon B with spin up and
spin down respectively.
We have used rotational invariance to obtain the last equality.

The polarization of a baryon from a polarized struck quark $q$
is then given by 

\beq 
P(q\uparrow \rightarrow B) = 
{{W(q\uparrow \rightarrow B\uparrow) - W(q\uparrow \rightarrow
B\downarrow)} 
\over 
{W(q\uparrow \rightarrow B\uparrow) + W(q\uparrow \rightarrow
B\downarrow)}} 
={{q\uparrow (B\uparrow)-q\downarrow (B\uparrow)}\over 
{q\uparrow (B) +q\downarrow (B)}}
\label{WWX}
\eeq

This can be rewritten in terms of the conventional notation for the
contribution of quark $q$ to the spin of baryon $B$, $\Delta q(B)$. 

\beq 
P(q\uparrow \rightarrow B) = 
{{\Delta q(B) } \over 
{q\uparrow (B) +q\downarrow (B)}}
\label{WW2}
\eeq

Expression (\ref{WW2}) illustrates the main problems in going beyond the
simple quark model to obtain the $\Lambda$ polarization. First one
needs to
know the contribution of the quark $q$ to the spin of the $\Lambda$.
This is not known directly and needs a symmetry or model to use the measured 
values for the proton to obtain the information on the $\Lambda$. A more 
serious difficulty is the normalization factor which is the total number of 
quarks of flavor $q$ in the $\Lambda$. This is trivial for the valence
quarks but is completely unknown for the sea quarks. 


We now consider the process in which a polarized struck quark $q$ with
spin projection $m=+1/2$ fragment into a $\Lambda$, $\Sigma^o$ or $\Sigma^*$.
In the  na\"{\i}ve quark model the lowest (uds) states with 
arbitrary spin couplings are linear combinations of $\Lambda$, $\Sigma^o$ and
$\Sigma^*$. The strange quark in the $\Lambda$ carries the full $\Lambda$ 
polarization and the nonstrange quarks are coupled to spin zero. 
The $\Sigma^o$ and $\Sigma^*$ both decay into a $\Lambda$ and 
we are interested
in the polarization of the $\Lambda$'s produced in all three ways.
In both the $\Sigma^o \rightarrow \Lambda \gamma$ and the 
$\Sigma^* \rightarrow \Lambda \pi$ decays, the nonstrange diquark emits the
boson in a J=1 state and changes its spin from one to zero, while the strange
quark is a spectator and its polarization is unchanged. Thus in the simple 
quark model the $\Lambda$ polarization is the same as the strange quark 
polarization for all three baryons. In the general case the polarization
of the \la from decay is still determined by the angular momentum couplings in
the decays and must therefore give the same result for any model.
The value of the polarization for any model is thus:
\beq
P_\Lambda ~ (B,M)= P_s (B,M) = 2M\cdot \Delta s (B)_{NQM} 
\label{QQ1}
\eeq
where $\Delta s (B)_{NQM}$ denotes the value of $\Delta s$ given by the 
na\"{\i}ve quark model for baryon $B$.
To obtain the value of the observed polarization it is necessary to average
the result (\ref{QQ1}) over all relevant $M$ states. Since all $M$ states of
the $uds$ system are expected to be produced equally, we need to find the 
relative probability $W(B,M)_q$ 
that a given $M$ state of baryon $B$ contains a quark $q$
with m=+1/2. This is simply given in terms of the contribution
of a quark of flavor $q$ to the total spin of baryon $B$.
\beq
W(B,M)_q = (1/2) + M\cdot \Delta q(B) 
\label{QQ2}
\eeq
Combining eqs. ( \ref{QQ1}) and ( \ref{QQ2}) and averaging over $M$ then gives
\bea
P_\Lambda (B)_q = P_s (B)_q &=& 
{{\sum_M 2M\cdot \Delta s (B)_{NQM} \cdot W(B,M)_q }\over{\sum_M W(B,M)_q}}= 
\nonumber\\
&=&{{\sum_M \Delta s(B)_{NQM} \cdot [M + 2M^2 \cdot \Delta q(B) ]}\over
{\sum_M (1/2) + M\cdot \Delta q(B) }}= 
\nonumber\\
&=& {{4 \sum_M M^2 }\over {2J+1 }} \cdot  \Delta s(B)_{NQM} \cdot \Delta q(B) 
\label{QQ3}
\eea
where we have noted that $\sum_M M = 0$ and $\sum_M (1/2) = (2J+1)/2$. 

\noindent
For the $\Lambda$ and $\Sigma^o$, with J=1/2 
\beq
{{4 \sum_M M^2 }\over {2J+1 }} = 1; ~ ~ ~ P_\Lambda ~ (B)_q = 
\Delta s(B)_{NQM} \cdot \Delta q(B) 
\label{QQ4}
\eeq
For the $ \Sigma^*$, with J = 3/2, 
\beq
{{4 \sum_M M^2 }\over {2J+1 }} = 5; ~ ~ ~ P_\Lambda ~
(\Sigma^*)_q =  5 \Delta s(\Sigma^*)_{NQM} \cdot \Delta q(\Sigma^*) 
\label{QQ5}
\eeq
The na\"{\i}ve quark model values of $\Delta q(B) $ are:
\beq
\Delta u (\Lambda) = \Delta d (\Lambda) = 0; ~ ~ ~ \Delta s (\Lambda) = 1
\label{QQ6}
\eeq
\beq
\Delta u ( \Sigma^o) = \Delta d (\Sigma^o) = 2/3; ~ ~ ~ \Delta s (\Sigma^o) = 
- 1/3
\label{QQ7}
\eeq
\beq
\Delta u (\Sigma^{*o}) = \Delta d (\Sigma^{*o}) = \Delta s (\Sigma^{*o}) = 1/3
\label{QQ8}
\eeq
We then obtain for a general model:

\beq
P_\Lambda   ~ ( \Sigma^*)_{u,d,s}  = (5/3) \cdot \Delta q(\Sigma^*)
\eeq
\beq
P_\Lambda  ~ (\Sigma^o)_{u,d,s}  = -(1/3)\cdot \Delta q(\Sigma^o)
\eeq
\beq
P_\Lambda  ~ (\Lambda)_{u,d,s}  = \Delta q(\Lambda)
\eeq

We can write the \la polarization as proportional to the polarization of the
struck quark $q$: $P_{\Lambda} = ~c \cdot P_q$. For the na\"{\i}ve quark model
which is expected to be correct for large $x$ \cite{aleph,opal},
we can use the values cited above. The proportionality coefficients
$c$ are summarized in table 1. In the hadronization process the struck quark 
may hadronize directly to a \la, in which cas it is also the ``Parent", or
it may hadronize to a Hyperon that decays to the \la and then that Hyperon
is the parent. It may also happen that a \la was produced without carrying
the struck quark at all and in this hadronization the \la is described in
the table as having a ``$q,\bar{q}$" parent.\\

\begin{table}[t]
\caption{ The coefficients $c$ in the equation $P_{\Lambda} = ~c \cdot P_q$
for q = u,d,s derived for the na\"{\i}ve quark model. The parent noted as 
$q,\bar{q}$ is a fragmentation in which the struck quark does not participate.}
  \centering
\vskip 1.0cm
  \begin{tabular}{|c|c|c|c|} \hline\hline
Parent $\setminus$ struck & u & d & s \\
\hline\hline
  &  &  &    \\
$q,\bar{q}$ & 0 & 0 & 0 \\
  &  &  &    \\
s & 0 & 0 & 1 \\
  &  &  &    \\
$\Sigma^0$ & $-\frac{2}{9}$ &  $-\frac{2}{9}$ &  $\frac{1}{9}$ \\
  &  &  &    \\
$\Sigma^*$ & $\frac{5}{9}$ & $\frac{5}{9}$ & $\frac{5}{9}$ \\
  &  &  &    \\
\hline\hline
\end{tabular}
\label{tab:qm}
\end{table}

An alternative approach is to assume that the spin structure of 
SU(3) octet hyperons
can be deduced from that of the proton as measured in polarized DIS
experiments.  Now baryons
consist of three valence quarks and a sea. In order to get the spin 
structure of the \la, $\Sigma^0$ and $\Sigma^*$ from that of the nucleon
we assume that the  quark wave functions for the hyperons are 
obtained from the nucleon wave functions by using SU(3) symmetry. 
The  $\Sigma^*$,
which is not member of the octet is treated separately. We follow the
notations of \cite{jaf} defining $\Delta Q = \Delta q + \Delta \bar{q}$.
SU(3) symmetry gives the following relations: 

\begin{equation}
\label{eq:su3}
\Delta U (p) = \Delta D (n) = \Delta U (\Sigma^+) = \Delta D (\Sigma^-) = 
\Delta S (\Xi^o) = \Delta S (\Xi^-)
\end{equation}
\begin{equation}
\Delta D (p) = \Delta U (n) = \Delta S (\Sigma^+) = \Delta S (\Sigma^-) 
= \Delta S (\Sigma^o) = \Delta U (\Xi^o) = \Delta D (\Xi^-) 
\end{equation}
\begin{equation}
\Delta S (p) = \Delta S (n) = \Delta D (\Sigma^+) = \Delta U (\Sigma^-) = 
\Delta D (\Xi^o) = \Delta U (\Xi^-)
\end{equation}
\begin{equation}
\Delta U (\Sigma^o) = \Delta D (\Sigma^o) = (1/2)\cdot 
[\Delta U (\Sigma^+) + \Delta D (\Sigma^+) ]
\end{equation}
\begin{equation}
\Delta Q (\Sigma^o) + \Delta Q (\Lambda) = (2/3)\cdot 
[\Delta U (n) + \Delta D (n) + \Delta S (n)] 
\label{eq:su3_5}
\end{equation}

These relations allow the values of $\Delta U ,~\Delta D$ and $\Delta S$
for all the octet baryons  to be obtained from the values for the proton.
The results are summarized in table \ref{tab:su3}\\

\begin{table}[t]
\caption{The contributions from quarks and antiquarks to the Baryon Spin 
Structures from Nucleon Data and SU(3) Symmetry. The 
values for the proton are taken from reference \protect\cite{slac}.}
  \centering
\vskip 1.0cm
  \begin{tabular}{|c|c|c|c|} \hline\hline
Baryon &  $\Delta U$ &  $\Delta D$ &  $\Delta S$ \\
\hline\hline
  &  &  &    \\
p & ~~0.83 $\pm$ 0.03 & $-0.43 ~\pm$ 0.03 & $-0.10 ~\pm$ 0.03 \\
n & $-0.43 ~\pm$ 0.03 & ~~0.83 $\pm$ 0.03 & $-0.10 ~\pm$ 0.03 \\
$\Sigma^+$ & ~~0.83 $\pm$ 0.03 & $-0.10 ~\pm$ 0.03 & $-0.43 ~\pm$ 0.03 \\
$\Sigma^-$ & $-0.10 ~\pm$ 0.03 & ~~0.83 $\pm$ 0.03 & $-0.43 ~\pm$ 0.03 \\
$\Xi^o$ & $-0.43 ~\pm$ 0.03 & $-0.10 ~\pm$ 0.03 & ~~0.83 $\pm$ 0.03 \\
$\Xi^-$  & $-0.10 ~\pm$ 0.03 & $-0.43 ~\pm$ 0.03 & ~~0.83 $\pm$ 0.03 \\
$\Sigma^o$ & ~~0.37 $\pm$ 0.03 & ~~0.37 $\pm$ 0.03 & $-0.43 ~\pm$ 0.03 \\
$\Lambda$ & $-0.17 ~\pm$ 0.03 & $-0.17 ~\pm$ 0.03 &  ~~0.63 $\pm$ 0.03 \\
  &  &  &    \\
\hline\hline
\end{tabular}
\label{tab:su3}
\end{table}

\noindent
\underline{\bf \la from struck quarks and from  
$\Sigma^0 \rightarrow \Lambda\gamma$.}
In DIS the produced \la or \lab is associated with the photon
being absorbed on a quark or antiquark, respectively. The struck quark
or antiquark then hadronizes into the \la or \lab (or another parent hyperon
which subsequently decays to a \la or \lab).
It is therefore necessary to have the separate contributions from the 
quarks and antiquarks to the spin structure functions. 
Following \cite{jaf} we calculate these values assuming that
the sea quark polarization distributions is the same in all the octet
members. We then make the calculations assuming that the sea is SU(3) flavor 
symmetric. For the \la this assumption means:
$\Delta s_N ~= ~\Delta \bar{s}_N ~= ~\Delta \bar{u}_{\Lambda}
 ~= ~\Delta \bar{d}_{\Lambda}  ~= ~\Delta \bar{s}_{\Lambda} ~= ~...$. In
particular $\Delta \bar{u}_{\Lambda} ~= ~\Delta \bar{s}_N ~=~ 
\frac{1}{2}\Delta S_N ~=~ -0.05 \pm 0.015$. Similarly for  $\Sigma^0$.
In order to get the contributions from the valence quarks only,
the sea $\Delta Q$ are assumed to be $-0.1 ~\pm$ 0.03. By
subtracting this value from the total $\Delta Q$ we obtain the value
of the valence $\Delta q$ (no antiquarks here). The resulting
values for the valence quarks are listed in table \ref{tab:qsu3}.
An analogous table would have antibaryons and antiquarks replacing
the baryons and the quarks.\\

\begin{table}[t]
\caption{The contributions from only quarks to the baryon spin 
structures assuming the sea is SU(3) symmetric}
  \centering
\vskip 1.0cm
  \begin{tabular}{|c|c|c|} \hline\hline
Distribution & all quarks &  valence quarks  \\
\hline\hline
  &  &   \\
$\Delta u_p$ & ~~0.88 $\pm$ 0.03 & ~~0.93 $\pm$ 0.03 \\
$\Delta d_p$ & $-0.38 ~\pm$ 0.03 & $-0.33 ~\pm$ 0.03 \\
$\Delta s_p$ & $-0.05 ~\pm$ 0.03 & ~0.0 $\pm$ 0.03 \\
$\Delta u_{\Lambda}$ & $-0.12 ~\pm$ 0.03 & $-0.07 ~\pm$ 0.04 \\
$\Delta d_{\Lambda}$ & $-0.12 ~\pm$ 0.03 & $-0.07 ~\pm$ 0.04  \\
$\Delta s_{\Lambda}$ & ~~0.68 $\pm$ 0.03 & ~~0.73 $\pm$ 0.04  \\
$\Delta u_{\Sigma^0}$ & ~~0.42 $\pm$ 0.03 & ~~0.47 $\pm$ 0.04  \\
$\Delta d_{\Sigma^0}$ & ~~0.42 $\pm$ 0.03 & ~~0.47 $\pm$ 0.04  \\
$\Delta s_{\Sigma^0}$ & $-0.38 ~\pm$ 0.03 & $-0.33 ~\pm$ 0.04  \\
$\Delta u_{\Sigma^*}$ & & ~~0.23 $\pm$ 0.02  \\
$\Delta d_{\Sigma^*}$ & & ~~0.23 $\pm$ 0.02  \\
$\Delta s_{\Sigma^*}$ & &~~0.33 $\pm$ 0.04   \\

  &  &   \\
\hline\hline
\end{tabular}
\label{tab:qsu3}
\end{table}

It is known experimentally that SU(3) symmetry is broken for the sea,
the polarized s-quark contribution being
suppressed by a factor of 2 compared with that of the non-strange quarks
\cite{ccfr}. However, the manner in which the spin
contributions are divided between valence and sea quarks is model
dependent. We follow the model in which the sea is assumed to be an SU(3)
singlet \cite{zvi12};  i.e. that it is not polarized in flavor space by
the valence quarks. In this case the sea is the same for all octet baryons
and the valence quarks satisfy SU(3) symmetry. Thus equations 
\ref{eq:su3} - \ref{eq:su3_5} also apply only to the valence quarks. In
this model the breaking of SU(3) symmetry by suppressing the strange quark
contribution to the sea does not affect the valence quarks and the sea is
still assumed to be the same for all octet baryons. This assumption is
justified in detail in ref. \cite{zvi12}. Consequently, the values listed 
in table \ref{tab:su3} will change but those of table \ref{tab:qsu3} will
not. For calculation of the polarization of \la hyperons produced in the
current fragmentation region we need only the values of $\Delta q$ for
valence quarks as listed in table \ref{tab:qsu3}. Thus the results are
not sensitive to SU(3) symmetry breaking.\\

The polarization of the \la and \lab is now computed according to eq.
\ref{WW2}.
For direct \la production the polarization will be equal to the
contribution 
of the struck quark as given in table \ref{tab:qsu3}. If the \la is
detected with a relatively large $x_F$ it is likely that the struck
quark is a valence quark in the \la resulting from its hadronization.
For example, if a
100\% polarized photon strikes out a s-quark, the quark will also be
100\% polarized. However, when it hadronizes into a \la there will be
contributions from the u- and d-quarks, the sea and relativistic
corrections that will cause the \la polarization to be only about 70\%. 
For \la from $\Sigma^0 \rightarrow \Lambda\gamma$ we compute the
polarization of the \la from eq. \ref{QQ4} and 
table \ref{tab:qsu3}.  Again, we write
$P_{\Lambda} = ~c \cdot P_q$ and the proportionality coefficients 
$c$ are summarized in table \ref{tab:su}.\\

\noindent
\underline{\bf \la from  $\Sigma^* \rightarrow \Lambda\pi$.}
In order to get the contributions from the valence quarks of the  $\Sigma^*$
to its spin we cannot use SU(3) symmetry because the  $\Sigma^*$
is not member of the octet. We  assume that the contribution from valence
quarks of the hyperons is proportional to the contributions
expected from the quark model. We therefore relate the quark model
expectations for the $\Sigma^0$ and  $\Sigma^*$ and use the values for the
former from table \ref{tab:qsu3} to calculate the values for the latter.
 For $\Sigma^0$ we have from the quark 
model:
$\Delta$u = $\Delta$d = $\frac{2}{3}$ and $\Delta$s = $-\frac{1}{3}$
(eq. \ref{QQ7}). For $\Sigma^*$ we have: 
$\Delta$u = $\Delta$d = $\Delta$s = $\frac{1}{3}$ (eq. \ref{QQ8}). This
gives us ratios of the contributions
from  $\Sigma^*$ to those of  $\Sigma^0$ to be: 0.5, 0.5 and $-1.0$
for $\Delta u, \Delta d$ and $\Delta s$, respectively. Using these ratios
and the values for $\Sigma^0$ listed in table \ref{tab:qsu3} we calculate
the values expected for $\Sigma^*$. We do not estimate
contributions from the sea in the $\Sigma^*$. 
The results are listed in table \ref{tab:qsu3}.
The polarization of \la from $\Sigma^* \rightarrow \Lambda\pi$ is
computed using eq. \ref{QQ5} and the values of $\Delta q(\Sigma^*)$ are 
taken from table \ref{tab:qsu3}. The proportionality coefficients $c$ in
$P_{\Lambda} = ~c \cdot P_q$  are summarized in table
\ref{tab:su}. 


\begin{table}[h]
\caption{ The coefficients $c$ in the equation 
$P_{\Lambda} = ~c \cdot P_ q$ for q = u,d,s derived using SU(3) symmetry
for valence quarks. The parent noted as $q,\bar{q}$ is a fragmentation in 
which the struck quark does not participate. }
  \centering
\vskip 1.0cm
  \begin{tabular}{|c|c|c|c|} \hline\hline
Parent $\setminus$ struck & u & d & s \\
\hline\hline
  &  &  &    \\
$q,\bar{q}$ & 0 & 0 & 0 \\
  &  &  &    \\
s & $-$0.07 & $-$0.07 & ~~0.73 \\
  &  &  &    \\
$\Sigma^0$ & $-$0.16 &  $-$0.16 &  ~~0.11 \\
  &  &  &    \\
$\Sigma^*$ & ~~0.38 & ~~0.38  & ~~0.55 \\
  &  &  &    \\
\hline\hline
\end{tabular}
\label{tab:su}
\end{table}

\noindent
As an illustration we present some predictions for the polarization of \la
and \lab produced in DIS. We use a Lund Monte Carlo simulation based on 
LEPTO generator to predict the number of \la and \lab produced either
directly or as decay products as a function of $x_F$. We then use the
values presented in tables \ref{tab:qm} and \ref{tab:su} to calculate the 
polarization predicted by the quark model and the SU(3)$_F$ symmetry
model, respectively. The results for \la are presented in figure 1
showing the contributions from direct production and the two decay modes
(Fig. 1a) and comparing the two models (Fig. 1b). The results depend on
the Monte Carlo structure such as the parton distributions and
hadronization parameters as well as the particular DIS kinematics. The 
results in Figure 1 were calculated for the conditions of Fermilab
experiment
E665 \cite{e665}. It also should be noted that all the calculations
presented in this work are justified only for the current fragmentation
region. Therefore they should not be reliable for low $x_F$ values.\\


\bibliographystyle{unsrt}

\newpage
\begin{figure}[t]
\epsfig{figure=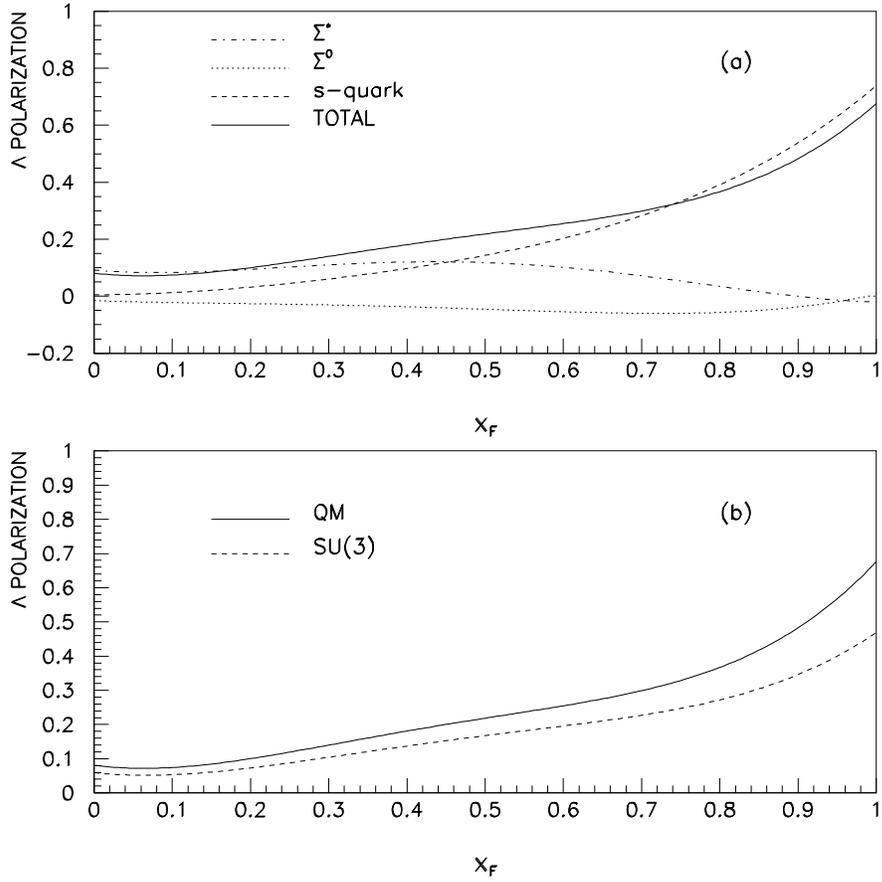,width=13cm}
\caption{Polarization of \la hyperons. (a) Contributions from direct
production (dashed line), from decays of $\Sigma^0$ (dotted line),
of $\Sigma^*$ (dash-dotted line) and the total
polarization (solid line). All are calculated using the  na\"{\i}ve quark
model.
(b) The $\Lambda$ polarization from all components calculated using the
 na\"{\i}ve quark model (solid line) and the SU(3) symmetry model (dashed
line).}
\label{calc}
\end{figure}

\end{document}